# A model for relative biological effectiveness of therapeutic proton beams based on a global fit of cell survival data

Ramin Abolfath[1,3,4*], Christopher R. Peeler[1], Mark Newpower[1], Lawrence Bronk[2], David Grosshans[2], and Radhe Mohan[1#]

Departments of Radiation Physics[1] and Radiation Oncology[2], The University of Texas MD Anderson Cancer Center, Houston, TX 77030; [3]Department of Therapeutic Radiology, Yale University School of Medicine, New Haven, CT 06520; [4]Department of Radiation Oncology, St. Vincent's Hospital, Bridgeport, CT 06606

Corresponding Authors:
* ramin1.abolfath@gmail.com / ramin.abolfath@yale.edu
# rmohan@mdanderson.org

**Abstract**

We introduce an approach for global fitting of the recently published high-throughput and high accuracy clonogenic cell-survival data for therapeutic scanned proton beams. Our fitting procedure accounts for the correlation between the cell-survival, the absorbed (physical) dose and the proton linear energy transfer (LET). The fitting polynomials and constraints have been constructed upon generalization of the microdosimetric kinetic model (gMKM) adapted to account for the low energy and high lineal-energy spectrum of the beam where the current radiobiological models may underestimate the reported relative biological effectiveness (RBE). The parameters $(\alpha, \beta)$ of the linear-quadratic (LQ) model calculated by the presented method reveal a smooth transition from low to high LETs which is an advantage of the current method over methods previously employed to fit the same clonogenic data. Finally, the presented approach provides insight into underlying microscopic mechanisms which, with future study, may help to elucidate radiobiological responses along the Bragg curve and resolve discrepancies between experimental data and current RBE models.

**Introduction**

One of the main advantages of using protons and other ions in radiation therapy is the Bragg peak and its sharp distal dose gradient. This enables high dose delivery to deep seated tumors [1].

While biological optimization is used for heavy ion therapy, for proton therapy a constant relative biological effectiveness (RBE) value of 1.1 continues to be employed clinically, despite experimental evidence that indicates a non-static RBE. A complete and robust model describing the cellular response and the biological effectiveness of protons along the Bragg curve would facilitate the biologic optimization of proton plans. However, uncertainties in the biological data at locations at or just distal to the Bragg-peak, where the charged particle LET is the highest, represent a particular challenge for the biological measurements and dose calculation in treatment planning systems [2-6].

In general, RBE depends on the radiation dose, the fractionation scheme, the biological endpoint, the radiation quality, dose rate, and the tissue and cell-type. Radiation quality includes the type of radiation, the energy and the linear energy transfer (LET) that describes the energy deposited from the beam into the irradiated material, per unit particle path length (in units of $keV/\mu m$). As a charged particle penetrates tissue, it slows down and its LET increases. Experimentally RBE reaches a maximum at a certain depth and then decreases [7].

A constant RBE of 1.1 has been recommended for clinical proton therapy [8, 9, 10, 11,

12], despite the fact that variable RBE values have long been found experimentally [13, 14, 15, 16, 17]. *In vitro* experimental data reveal that the RBE depends on dose and the proton energy spectrum at the point of measurement [13, 14]. On the other hand, *in vivo* biological responses present a more complex expression showing smaller variations in RBE values for shallow depths. More recent studies have shown a significant increase of RBE at the end of the proton range [15, 16, 17, 18].

Given these laboratory results, it is likely necessary to develop a method to predict the biological effects of proton beams for given physical and geometric characteristics. One of the first attempts at modeling RBE based on track structure theory of heavy ions and neutrons was proposed by Butts and Kraft [19, 20, 21]. Implementation of such models in clinical applications requires physical input parameters such as, local particle energy spectrum and, biological input parameters, such as cell nuclear size, end damage response, etc. Currently, access to such parameters is lacking due to the computational complexities required to obtain them.

Scholz, Krämer, and Kraft [22-25] implemented track structure models into particle radiation therapy that led to a model known as local effect model (LEM). Subsequently Paganetti [26] studied amorphous track structure models in proton radiation therapy and showed that the model needed a critical adjustment of the parameters to improve prediction. Recently, Elsasser et al. [27] proposed a modified local effect model which correlated spatial radiation damage and achieved good agreement with experimental data for different cell lines and several particle types, namely for protons and carbon ions.

First principle approaches have also been tried in many studies [28, 29, 30, 36, 37]. Neary discussed a linear dependence of chromosome aberrations with LET [36]. Hawkins derived a complex exponential relationship between $\alpha$ and lineal-energy and LET using microdosimetric kinetic modeling for several cell lines [28, 29]. The Wilkens and Oelfke linear model [30] predicted the experimental data for LET values up to 30 $keV/\mu m$ [30] well but was unable to predict the decrease in RBE with further increases of LET. Subsequent refinements of this model were reported in Refs. [31] and [32]. The repair-misrepair fixation (RMF) model of chromosome aberration that is one of the bases of our present model and computational analysis was developed in Refs. [33-35].

The Particle Irradiation Data Ensemble (PIDE) is an effort by Gesellschaft für Schwerionenforschung (GSI) to organize hundreds of radiation biology experiments in one dataset to facilitate evaluation of RBE models [39]. Based on their analysis of over 800 cell survival experiments, they conclude that RBE depends on particle species and the $\alpha/\beta$ ratio and that using LET alone is insufficient to predict RBE. In agreement with this assessment, the experimental data gathered in PIDE is a good collection to examine the fitting method introduced in the current study. We postpone to present this study to our future works and publications. In addition, Mohan et al. [38] recently provided a comparison of the RBE data from Guan *et al.* for H460 non-small cell lung cancer cells to the RBE predicted by a number of models, including those from Wilkens and Oelfke [30], Wedenberg *et al.* [32], Carabe-Fernandez *et al.* [40], McNamara *et al.* [41], Chen and Ahmad [42], Carlson *et al.* [33,51], and Frese *et al.* [35]. All of the models predicted lower RBE for the higher LET values used in the experiments. In this work we therefore focus on constructing a model for experimental data recently published by Guan *et al.* [37] as existing models do not adequately explain these data.

Our goal is to provide a theoretical framework in resolving the discrepancies between such experimental data and theoretical models. As detailed previously, the design of a graded solid water compensator (jig) allows irradiation of cells by a mono-energetic scanning beam of protons at specific depths. The absorbed dose and the proton LET were calculated using the Geant4 Monte Carlo toolkit [37]. Subsequent high-throughput, automated clonogenic survival assays were performed to spatially map the biological effectiveness of scanned proton beams with high accuracy. This method aims to reduce uncertainties in the biological data.

In this study we developed an approach for global fitting of the recently published high-throughput and high accuracy clonogenic cell-survival data by Guan et al.

### 2 Methods

We start by describing our phenomenological model, which is based on an extension of the microdosimetric kinetic model (MKM) originally proposed by Hawkins [28, 29]. We refer to our extension as generalized MKM (gMKM). In the following we elucidate the basis of gMKM and discuss in detail the differences with MKM.

In gMKM we first incorporate the effects of particle energy deposition spatial patterns calculated for single-tracks with nanometer spatial resolution. The initial energy of the proton beam and the geometry were consistent with the experimental beam line used in Ref. [37]. The proton energy ranges from 80 MeV to as low as 100 keV. The choice in low energy cut-off is based on the range of protons with 100 keV of energy, which is less than the diameter of a cell nucleus, e.g., 5 $\mu m$ (typical diameter range is within 3-7 $\mu m$ depending on several factors such as cell type, cell cycle, adherence, etc.). To this end, we used a combination of Geant4 and Geant4-DNA Monte Carlo toolkits (version 10.2) to simulate the track-structures in macroscopic and microscopic scales in water equivalent materials [43].

We then analyzed the energy deposition distributions of protons and assessed if the Poisson distribution that is the central assumption in MKM effectively described the track structure. To this end, we randomly selected locations along the beam axis to score energy deposition in 5 $\mu m$-diameter spherical volumes representing cell nuclei. The dimensions of these volumes were consistent with the dimensions of the cell nucleus used in the clonogenic assays in Ref. [37].

With these two steps, we systematically obtained the distribution of ionizations in geometrical structures resembling the cell nucleus as a function of their position relative to the source, proton energy, and LET along the radiation track.

As shown in Figure 1a, in locations near the beam entrance where the energy of the proton is relatively high, i.e. 80 MeV, and therefore the LET is low, the local density of ionization within the cell nucleus is sparse and can be fitted by a Gaussian and/or Poisson distribution. The spectrum of energy loss and lineal-energies as a function of depth are plotted in Figure 2. The Gaussian / Poissonian approximation of energy-loss spectral density holds for a wide range of depths, from the entrance to locations proximal to the Bragg peak where the protons slow down to energies of approximately 15 MeV and $LET \approx 10\ keV/\mu m$. This result can be seen in Figure 1c where the calculated energy deposition and LET as a function of depth using Geant4 for a circular shape source of proton consists of a Gaussian energy broadening.

Beyond $LET \approx 10\ keV/\mu m$, as shown in Figure 1b, the atomistic excitations and ionizations undergo a transition to a highly compact track structure. This transition in energy loss spectrum to a clearly visible Landau distribution [57-58] occurs around $d = 52\ mm$ for the 80 MeV beam as shown in Figure 2. Therefore, above $LET \approx 10\ keV/\mu m$, the deviation from a Poisson distribution appears to be significant.

In the Appendix we develop an analytical model based on Neyman's distribution of type A (a compound Poisson distribution) [59-60] for illustration of the effects due to deviations from a Poisson distribution. In the following sections, however, we present numerical results of more sophisticated Monte Carlo calculations. In particular in Figure 2, we present a series of lineal energy spectra, fitted to the Landau distribution [57,58] for a beam of protons with 80 MeV of energy. The spectrum evolves as an increasing function of depth (depicted in clockwise order). An abrupt transition in the lineal energy spectrum is seen at 5.2 cm depth where the energy deposition coincides with the Bragg peak (see Figure 1c) and $LET \approx 10\ keV/\mu m$. The transition in the lineal energy spectrum corresponds to a similar transition in biological responses. As previously demonstrated, over a wide range of depths (highlighted by a square in the energy deposition curve in the left) LET is low and the spectrum can be approximated by a sharp and symmetric Gaussian / Poissonian distribution. At high LET's the Gaussian symmetry in lineal-energy distribution around its statistical mean value gradually breaks down. According to this analysis, the non-linearity in RBE stems from this symmetry breaking in the lineal energy spectrum. Moreover, we attribute a decrease in RBE beyond very high LET's, a phenomenon known as the over-kill effect in the literature [7], to the widening and sparsity of the lineal energy distribution (see for example the spectrum at $d = 55$ mm) and lack of beam strength due to a sharp drop in total energy deposition and particle fluence. These observations have been transpired to our practical approach in the fitting procedure where two kinds of linear and non-linear polynomials were converged and fitted to the cell survival data below and beyond $LET \approx 10 \; keV/\mu m$.

We now turn to highlight the mathematical framework we employed in fitting the experimental data using LET power series in $\alpha$ and $\beta$. We refer the interested readers to the mathematical details in the Appendix.

The results obtained from experiments [37] as well as simulations [35] confirm that as LET increases, $\alpha$ and $\beta$ evolve from linear to non-linear functions of LET. Similar to MKM, in gMKM we calculated the cell survival-fraction by integrating over the solutions of the repair-misrepair-fixation (RMF) model [33, 45, 46]. The RMF model links radiation double strand break (DSB) induction and the processes that lead to exchange-type chromosome aberrations and cell death. The linear-quadratic (LQ) survival model has been shown to be an approximate time-integrated solution to the RMF model that accounts for incorrect chromosome end joining. More accurate consideration of the RMF model by perturbative linearization of the non-linear solutions leads to the cell survival-fraction, $SF$, represented as a power series of the statistical moments of the specific energy $z$ in a cell-nucleus domain

$$-\ln(SF) = a_1 \bar{z} + a_2 \overline{z^2} + a_3 \overline{z^3} + \cdots. \quad (1)$$

Note that the higher order terms beyond the MKM LQ-model in Eq. (1) are from a power expansion of RMF solutions. Such non-linearity in SF have been examined before to analyze lack of symmetry in split-dose survival of mouse jejunal crypt cells where total dose was fixed but asymmetric combinations of the dose were used, i.e., large following small vs. small following large dose fractions [48]. In Eq.(1), $a_1, a_2, \ldots$ are expansion coefficients and $\bar{z}, \overline{z^2}, \ldots$ are the ensemble averaged moments of the specific energy. Note that $a_1, a_2, \ldots$ are combinations of the DNA-DSB induction strength constants in the RMF model and the pair-wise repair-misrepair chromosome end-joining rate coefficients.

To perform the averaging in Eq.(1) explicitly, the distribution function of these random variables must be known. In MKM, the Poisson distribution is assumed to govern the distribution of ionization events. Based on our analysis, this is a plausible assumption in the low LET limit, however, a correction to the Poisson distribution is warranted for high LETs. We therefore expand the non-Poissonian distribution function of ions in the vicinity of the Poisson distribution. Following algebraic manipulation, an expansion around the MKM mean specific energy $\bar{z}_{MKM}$ in Eq. (1) can be found. Using the macroscopic dose $D \equiv \bar{z}$ and dose-averaged lineal-energy $y_d$ or equivalent $L$ (simply LET that is a linear function of $y_d$) as two independent variables in this equation, we find

$$-\ln(SF) = \sum_{i=1}^{n} \sum_{j=1}^{i} b_{i,j} L^{i-j} D^j. \quad (2)$$

As described in the Appendix, derivation of Eq. (2) involves expansion of moments $\overline{z^2}, \overline{z^3}, \ldots$ of a non-Poissonian distribution, in power series of Poissonian or MKM specific energies, $\bar{z}_{MKM}, \bar{z}^2_{MKM}, \bar{z}^3_{MKM}, \ldots$, e.g., $\bar{z} = \sum_{k=1} c_k \overline{z^k}_{MKM}$; $\overline{z^2} = \sum_{k=2} c_k \overline{z^k}_{MKM}$; $\overline{z^3} = \sum_{k=3} c_k \overline{z^k}_{MKM}$; ... Thus the expansion coefficients $b_{i,j}$ are linear combinations of $a_1, a_2, \ldots$ and $c_1, c_2, \ldots$.

Recalling $\alpha$ and $\beta$ in the LQ cell-survival model, $-\ln(SF) = \alpha D + \beta D^2$, and truncating Eq. (2) up to the quadratic term in $D$ gives

and
$$\alpha = \sum_{k=1}^{n} b_{k,1} L^{k-1}, \qquad (3)$$

$$\beta = \sum_{k=1}^{n-1} b_{k+1,2} L^{k-1}, \qquad (4)$$

with $n \geq 2$. In Eqs. (3) and (4) $\alpha$ and $\beta$ are polynomials of LET with orders $n-1$ and $n-2$ respectively, thus the formula for $SF$ contains $2n-1$ fitting parameters. For clarification of the compact notations used in Eqs. (3) and (4), let us consider a class of polynomials truncated at $n=4$, hence $\alpha = b_{1,1} + b_{2,1}L + b_{3,1}L^2 + b_{4,1}L^3$ and $\beta = b_{2,2} + b_{3,2}L + b_{4,2}L^2$. A similar approach with slightly different representation of the RMF model as introduced by Carlson *et al.* [33] provides a more elaborate interpretation of the polynomials in the above equations. According to our unpublished studies, $n = 2, 3, 4, \ldots$ in $\alpha$ and $\beta$ in Eqs. (3) and (4) represent the levels of chromosomal aberration complexities such as the binary, ternary, quaternary, etc. chromosome recombination. We postpone the details of the higher order (beyond binary) chromosome recombination leading from Neyman's distribution of type A [59,60] to our future publications.

In practice we aim to find an optimal $n$ that allows the best fitting of the experimental data with minimal number of parameters. Thus, we need to calculate a minimal $n$. The lowest possible $n$ in Eqs. (3) and (4) corresponds to $n = 2$. This essentially corresponds to the linear $\alpha$ and constant $\beta$ as function of LET in MKM. This model contains three fitting parameters.

We now turn to discuss the results of the fitting procedure for the clonogenic cell survival experimental data. We use global fitting of $SF$ in a 3D parameter space using two independent variables $D$ and $L$. Eq. (2) provides specific forms of polynomials that we may use to fit the data.

### 3 Results and Discussion

In the current study we developed an approach for global fitting of the recently published high-throughput and high accuracy clonogenic cell-survival data.

A relation between cell survival and LET was extracted in a subsequent step after the calculation of the linear-quadratic cell survival parameters $\alpha$, $\beta$, $\alpha/\beta$ and RBE. The fitting was performed for each individual survival curve with a specific average LET. This method of fitting individually fits cell survival curves with measured data for each LET value separately, which would have led to high degree of variability in alpha and beta parameters of the LQ model. Instead, we carry out a "global" fit with the measured data. This process reduces the overall uncertainty. As a result, the dependence of cell-survival data on LET exhibits fluctuations, and it would be challenging to fit and quantitatively interpret the results into an RBE model. The source of numerical noise in fitted data largely stems from uncertainties in survival data. In practice, the LET dependence of cell-survival data may vary among different fitting procedures.

The data shows that the measured biologic effects are substantially greater than in most previous reports. It is characterized by a non-linear RBE as a function of LET near and beyond the Bragg peak. The calculated RBE is characterized by high sensitivity to small variations in LET distal to the Bragg peak, where a small uncertainty in the position of the cells may result in significant change in LET.

It is therefore crucial to search for appropriate fitting procedures and algorithms to be able to enhance the quality of the calculated RBE and the interpretation of the data. There are several models in the literature predicting regular and smooth dependence of RBE parameters on LET including the microdosimetric kinetic model (MKM), local effect model (LEM), and Monte Carlo damage simulation (MCDS). In this work, we explore more reliable fitting procedures which take into account a correlation among the SF curves. Our method is based on a three-parameter global fitting. We develop an optimization procedure that allows fitting a 2D surface in a 3D parameter space spanned by dose, LET and cell surviving fraction (SF).

An extension of this approach for particles heavier than protons is currently under investigation and the goal for such studies is to generate data needed to optimize treatment plans incorporating variable RBE.

A comparison between two schemes in global fitting using non-linear and linear polynomials are shown in Figures (3) and (4). In the linear polynomials fitting procedure, shown in Figure (4), we considered $\alpha = \alpha_0 + \alpha_1 LET;\ \beta = \beta_0$ and obtained an optimal surface in the three-dimensions spanned by dose, LET and SF that fit the experimental points whereas in the nonlinear approach shown in Figure (3), we used Eqs. (3) and (4). Clearly the linear fitting procedure can not cover the high LET data points. This is in contrast with the non-linear fitting. The error bar shown in these figures is extracted from the experimentally reported values in Ref.[37] and transformed to logarithmic scale using STD in log scale equivalent of STD/SF. Hence as we show the points with lower SF, the magnitude of error bars becomes more significant.

In the non-linear fitting procedure we further divided LET into two domains of high and low as discussed in the Methods section. For low LET, a linear $\alpha$ and constant $\beta$ which corresponds to choosing $n = 2$ in Eq. (2) can satisfactorily fit the data. The result of fitting for low LETs are shown in the left panel of Figure 5 (a and c).

For the completeness of our discussion, we reemphasize the expected transition from linear to non-linear biological responses along the energy deposition curve where LET increases monotonically. To characterize the threshold between "low" and "high" LETs, we calculated the lineal energy distribution function and used it as a metric to quantify the boundary between these two limits. In Figure 2 we summarized the evolution of the lineal energy spectrum for a proton pencil beam with initial energy 80 MeV as a function of depth and LET in water. As discussed earlier, a transition from sharply peaked and approximately symmetric distribution in low LETs to highly asymmetric Landau-type distribution takes place. This transition happens in the vicinity of the Bragg peak where LET is approximately $10\ keV/\mu m$ (see also Figure 1c). We therefore consider this value of LET as a border in fitting the cell survival data that divides the domains of the low and high LETs for protons. Thus we consider $LET\ \approx 10\ keV/\mu m$ in the global fitting of cell survival as the value below which a linear RBE model is used whereas above $LET\ \approx 10\ keV/\mu m$ a non-linear RBE model is implemented.

It is important to mention an abrupt transition in the lineal energy distribution may be attributed to a similar characteristic transition in the measurable biological quantities. For example the measured $\alpha$ and $\beta$ are the result of superposition-convolution of the corresponding parameters with spectrum of lineal energies, i.e., $\alpha(L) = \int_0^\infty \tilde{\alpha}(y)S(y)dy$, $\beta(L) = \int_0^\infty \tilde{\beta}(y)S(y)dy$ where $L = \int_0^\infty yS(y)dy / \int_0^\infty S(y)dy$, $S(y)$ is the lineal energy distribution function, and $\tilde{\alpha}$ and $\tilde{\beta}$ are theoretically defined for a single value of lineal energy, $y$.

The numerical values of the fitting parameters, presented in Table 1, were calculated within an in house developed Matlab code designed for an optimization algorithm in searching the best 2D surface fitted to a data set in 3D space. An iterative procedure was employed to minimize the chi-square value to obtain the optimal parameter values and performing nonlinear curve fitting. The optimization core of our approach uses the implementation of Levenberg-Marquardt-Fletcher algorithm developed for nonlinear least squares fitting problems.

As seen in this table we obtain a satisfactory convergence with $n = 5$ and $n = 6$ for H460 and H1437 cell lines. The rationale for selecting large values of $n$ stems from numerical methods in optimization and convergence of the self-consistent RBE solution in surface fitting procedure. The convergence of the power series can be seen from the numerical values presented in the table. By increasing the powers in the expansion, the numerical values of the coefficients will change, however, if the change in RBE resides within a convergence domain (chosen by the user) the optimization stops. Although in our approach we search for optimal surfaces to minimize chi-square values iteratively, it is known that additional metrics are needed to evaluate the goodness of the fit. We therefore report the calculated R-square, also known as coefficient of determination (COD) in Table 1. As shown in this table, the numerical values of R-square are close to 1. The closer the fit is to the data points, the closer R-square will be to the value of 1. It is also known that a larger value of R-square does not necessarily mean a better fit because of the degrees of freedom that can also affect the values. Thus we report the adjusted R-square values to account for the degrees of freedom.

To determine an optimal polynomial in the fitting process, we consider a series of polynomials corresponding to $n = 2, 3, 4, \ldots$ and perform optimization for each individual polynomial separately. A comparison based on chi-square and RBE is performed for polynomials with orders $n$ to $n + 1$. If the difference between two chi-square's (and RBE's) in steps $n$ and $n + 1$ is lower than a threshold value, e.g., $|\chi^2(n+1) - \chi^2(n)| < \delta$, the optimization stops and chi-square and RBE corresponding to step $n + 1$ is reported. In low LET's this procedure converges to a linear polynomial (as the zeros in Table 1 shows the contribution of higher order power series are negligible).

To examine the stability of the converged solutions we consider an alternative approach in matching the low and high LET surfaces at the boundary where $LET = 10$ keV/$\mu$m. To handle the discontinuity and obtain potentially smoother curves at the "transition point" where low LET data meet the high LET data, we include series of points below (and above) 10 keV/$\mu$m to high (and low) LET and perform the fitting procedure and calculate the optimum high (low) LET surfaces. We then remove these points located in the vicinity of the boundaries and compare the resulting surfaces with a simple piece-wise matching of low and high LET data at the boundaries as shown in Figures 7, 8 and 9. Comparison between RBE's, alpha's and beta's resulted in no significant difference.

In Figure 6 we show the same plots as in Figure 5 except the 3D surface is projected onto the SF-dose plane. The first few curves from the top in Figure 6 show the results of the $n = 2$ fitting for LETs less than 10 $keV/\mu m$. In Figure 7, , $\beta$, and RBE as a function of LET calculated from the fitted surfaces are plotted. For low LET, the linear dependence of $\alpha$, $\beta$, and RBE is consistent with the fitting procedures introduced in other publications including Hawkins [28, 29], Wilkens and Oelfke [30] and Steinsträter *et al.* [44]. $\alpha$

Recently McNamara et al. [41] fitted the present experimental data using a similar low-LET methodology [25]. The result of their calculation shows that present models may underestimate the high LET region of the LET-RBE relationship. This issue has been resolved within the presented methodology of gMKM. Using the fitting approach of the constructed framework, the last four red curves in Figure 6 (left panel), as well as in Figure 7, exhibit a reasonable fit to the high LET data. In this case, polynomials of order $n = 6$ were used for $\alpha$ and $\beta$.

In the right panel of Figure 6 we show the result of similar 3D global fitting using linear MKM polynomials, e.g., $\alpha = \alpha_0 + \alpha_1 L$ and a constant $\beta$. As shown in this figure, the low LET data can be fitted well by a linear $\alpha$ and a constant $\beta$, same as the left panel of Figure 6. However, for high LET the difference between the fitted curves and the experimental data is significant. Such difference can be reduced by using gMKM non-linear polynomials for $\alpha$ and $\beta$ as we used for the fitting of high-LET data in the left panel for Figure 6.

To illustrate the advantage of the present method, we compare the results from global 3D fitting with the conventional fitting procedure based on the same experimental data published recently by Guan et al. [37].

In the conventional fitting procedure, each survival curve with a specific LET was fitted individually without considering correlations among them. Hence, it may lead to irregular forms of $\alpha$ and $\beta$ as functions of LET. This method used to fit the data in Ref. [37], is inconsistent with the experimental design that was intended to correlate the cell-survival data of different LETs along the Bragg curve. Figures 8 and 9 show comparison between the conventional and 3D global fitting methods.

The presented fitting approach prevents the data from uncharacteristic fluctuations as shown in Figures 9 (circles). Instead it predicts a monotonic trend in $\alpha/\beta$ for all LETs [46]. Thus, the reported RBE's in Ref. [37] corresponding to fluctuations in $\alpha/\beta$, as shown in Figure 8, can be interpreted as an artifact associated with numerical instabilities in conventional fitting methods. This is consistent with a simplistic cell-inactivation target theory as single-track lethal events becomes more common due to increasing LET, thus the $\alpha/\beta$ ratio increases. This is consistent with a simplistic cell-inactivation target theory as single-track lethal events becomes more common due to increasing LET, thus the $\alpha/\beta$ ratio increases.

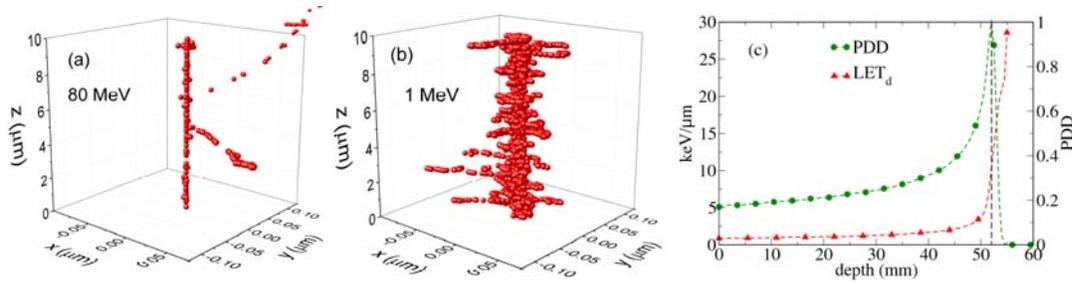

Figure 1. Two ionization-tracks of 80 MeV (a) and 1 MeV (b) protons obtained from the Geant4-DNA Monte Carlo simulation in water with initial points at x = y = z = 0 and directions parallel to z axis. The number of events in 1 MeV proton is one order of magnitude greater than 80 MeV. Such compact ionization events in low energies clearly justify a deviation from Poisson distribution that may describe the ionization in 80 MeV. The relative energy deposition along Bragg curve and LET for a beam of proton with 80 MeV is shown in (c). The line indicates the position of Bragg peak at approximate 5.2 cm depth corresponding with $LET \approx 10 keV/\mu m$.

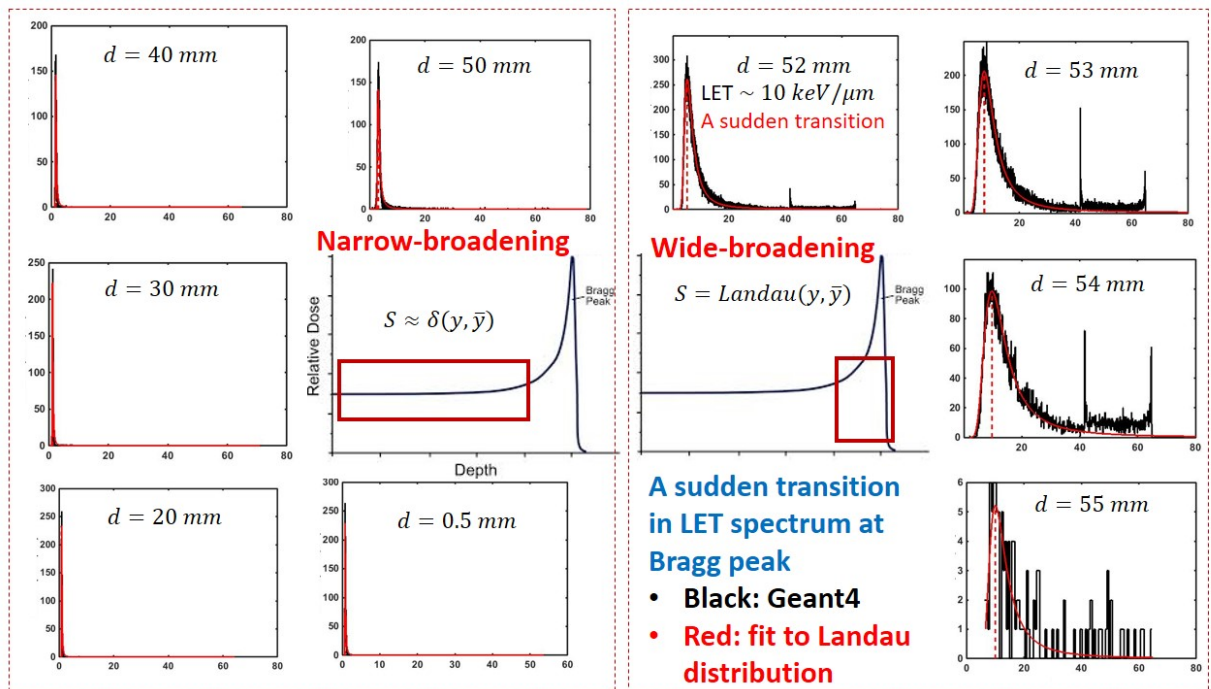

Figure 2. Series of the lineal energy spectrum and the fitting to Landau distribution is shown for a beam of proton with energy 80 MeV as an increasing function of depth in clockwise order. The x-axis and y-axis represent the lineal energy in $keV/\mu m$ and the histogram of lineal energy in

$(keV/\mu m)^{-1}$. An abrupt transition in shape of the lineal energy distribution function from sharply symmetric to wide and asymmetric form is seen at 5.2 cm depth where $LET \approx 10\ keV/\mu m$ (see Figure 1c) and LET exceeds $10\ keV/\mu m$.

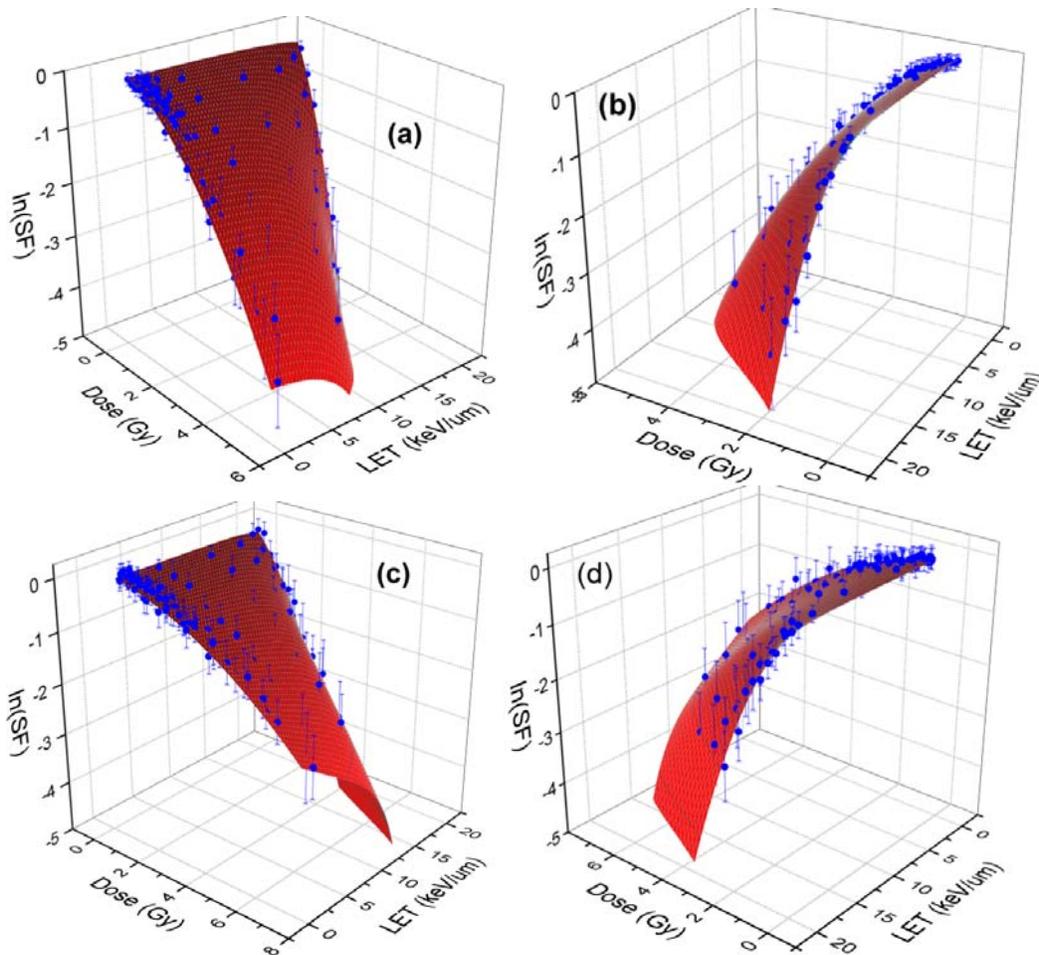

Figure 3: Experimental data of clonogenic survival fraction (SF) corresponding to H460 (a and b) and H1437 (c and d) cells are shown in blue dots with experimental error bars in logarithmic scale. The surfaces are fitted to the experimental data using global 3D optimization using a non-linear LET model as presented in the text. The left and right figures show the views from low and high LETs.

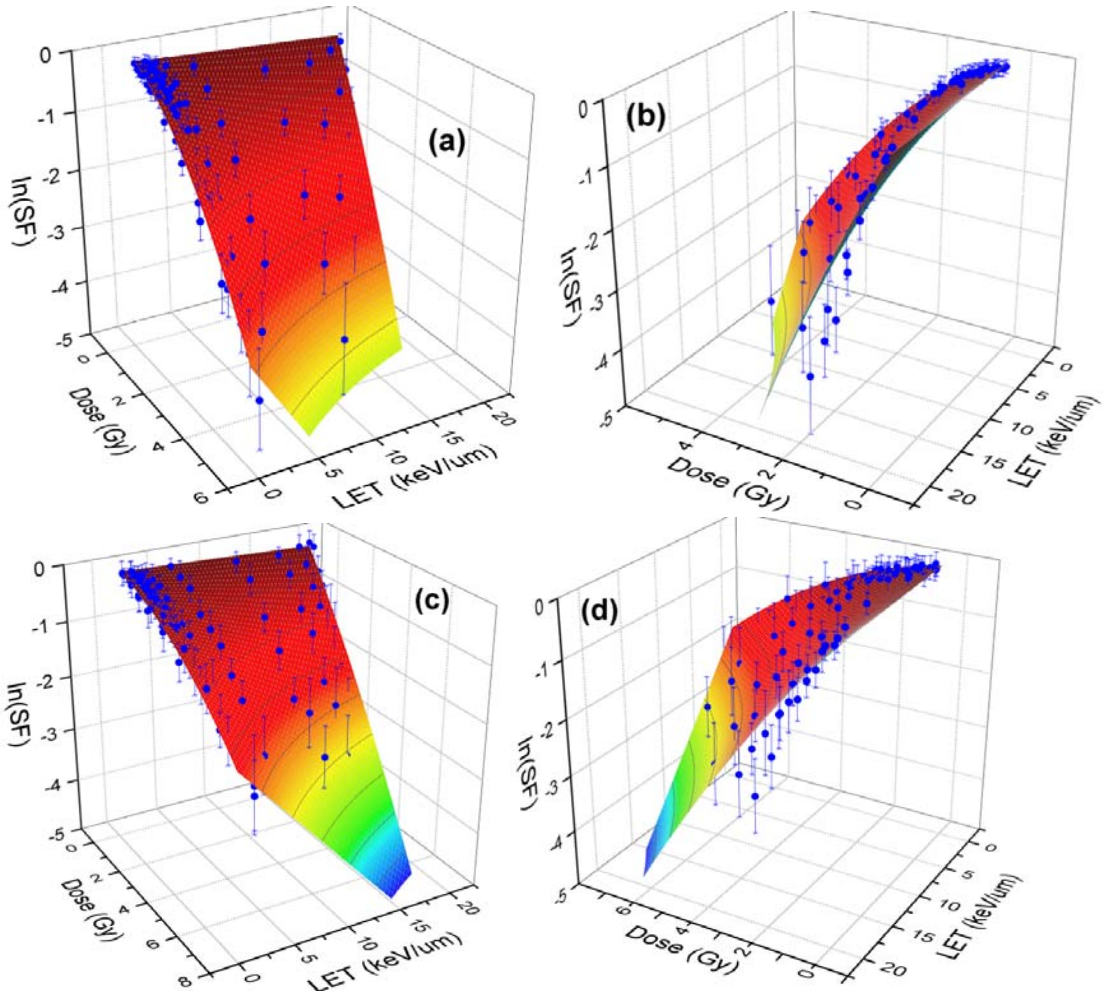

Figure 4: Shown same experimental data of clonogenic survival fraction (SF) for H460 (a and b) and H1437 (c and d) cells as in Figure 3 with a difference that a linear LET model corresponding to $\alpha = \alpha_0 + \alpha_1 L$ and $\beta = \beta_0$ have been used to obtain a globally optimal surface fitted to the experimental data. Similar to Figure 3, the left and right figures show the views from low and high LETs.

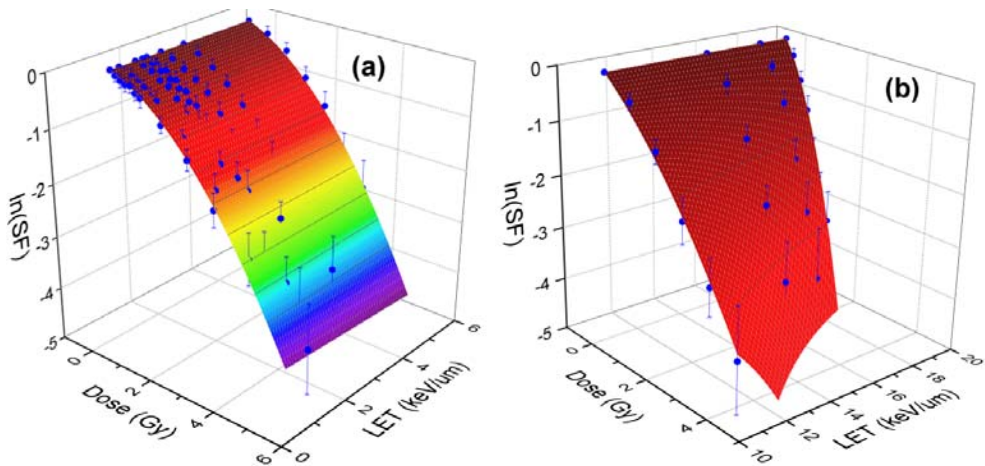

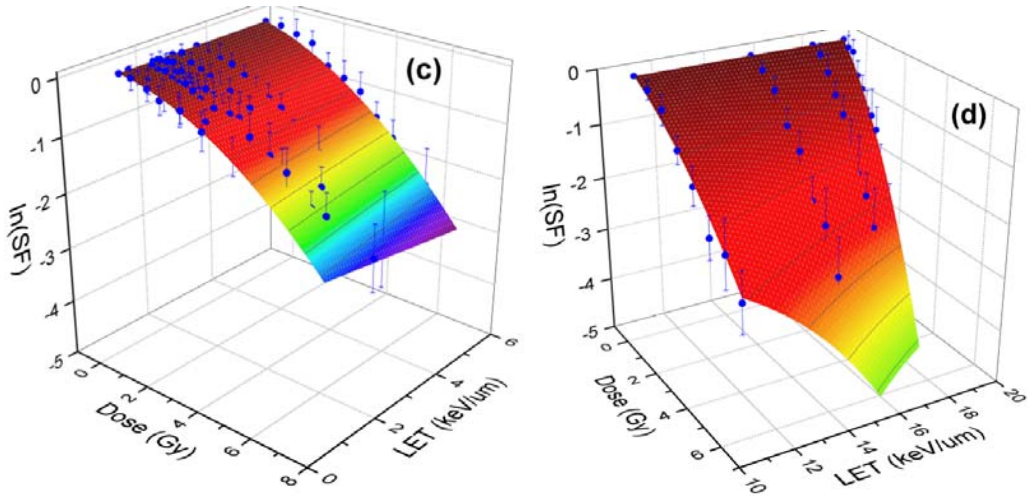

Figure 5: Shown H460 (a and b) and H1437 (c and d) cell survival surfaces in low (a and c) and high (b and d) LETs. As discussed in the text, $LET \approx 10 keV/\mu m$ separates the low LET from high LET domain for 80 MeV pencil beam of proton. The gMKM polynomials have been used in a 3D global fitting approach to obtain the best-fitted surface to the experimental data. The numerical values of the fitting parameters are in linear (a and c) and non-linear (b and d) domains are presented in Table 1.

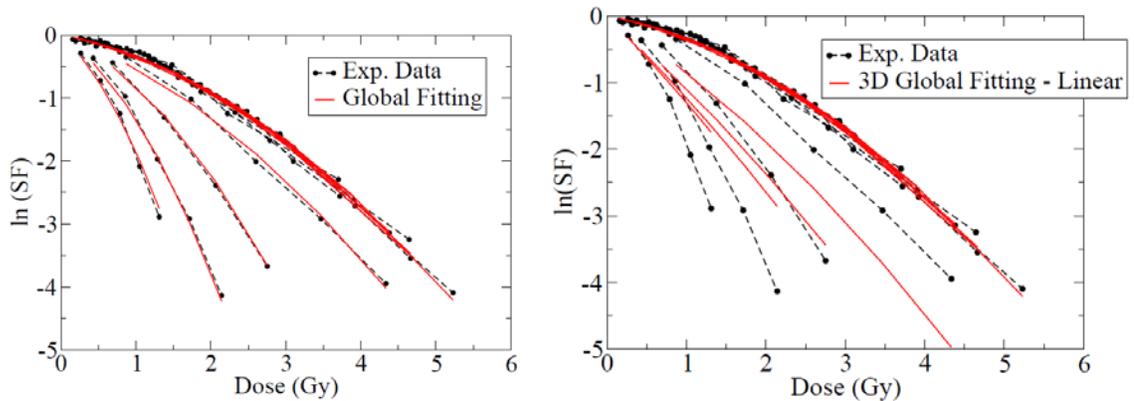

Figure 6: 2D representation of the survival fraction shown in Figure 3 (H460 cells) using non-linear gMKM fitting polynomials (left panel) and linear MKM polynomials in Figure 4 (right panel). The black dotted lines represent the experimental data and the red lines are the fitted surfaces projected to the SF-Dose plane.

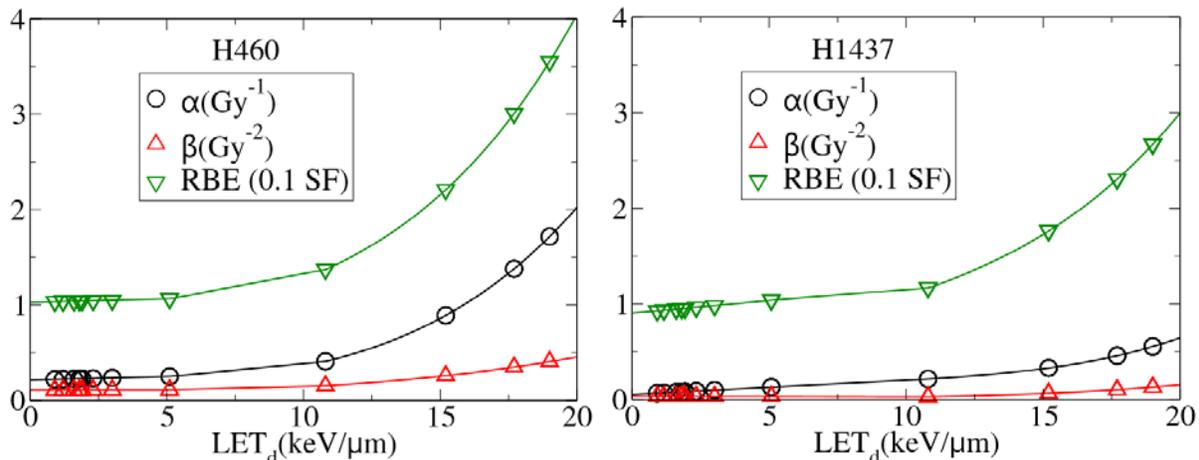

Figure 7: $\alpha$, $\beta$, and RBE as a function of LET calculated analytically from $SF$ equations obtained by the 3D global fitting of clonogenic cell-survival data as shown in Figures 2 and 3.

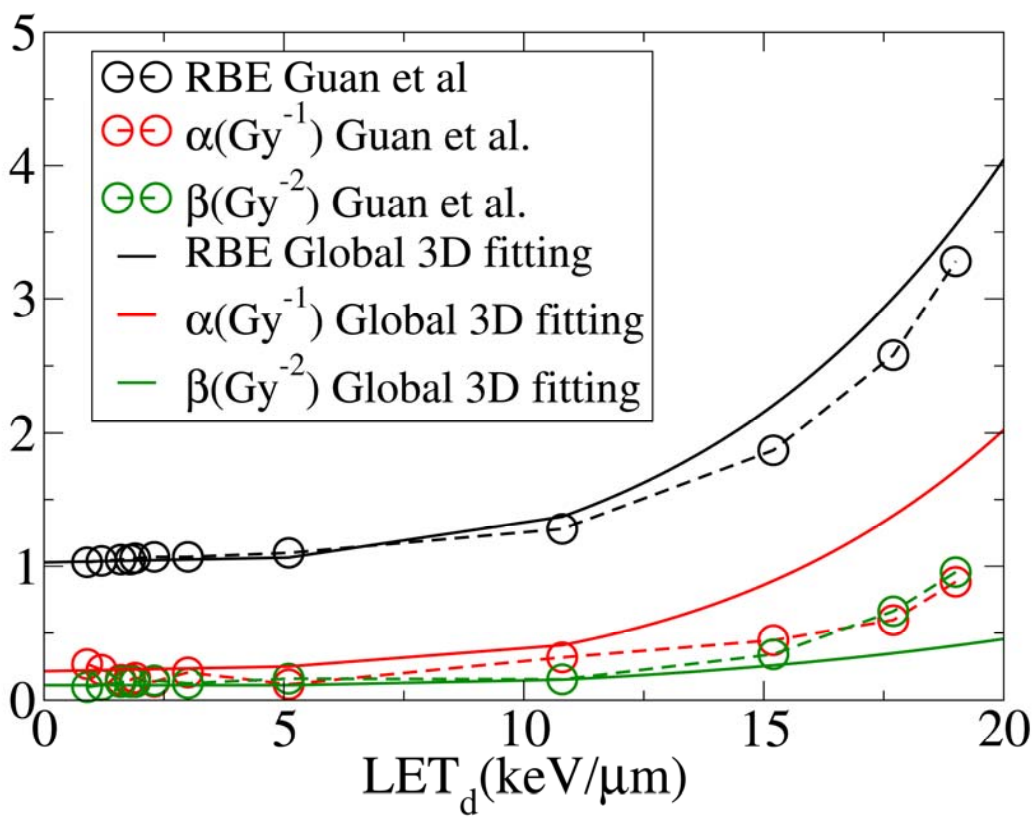

Figure 8: $\alpha$, $\beta$, and RBE as a function of LET calculated by the conventional method and the presented 3D global fitting method. Although the difference in RBE between the two methods is negligible, the $\alpha$ and $\beta$ are significantly different (H460 cells).

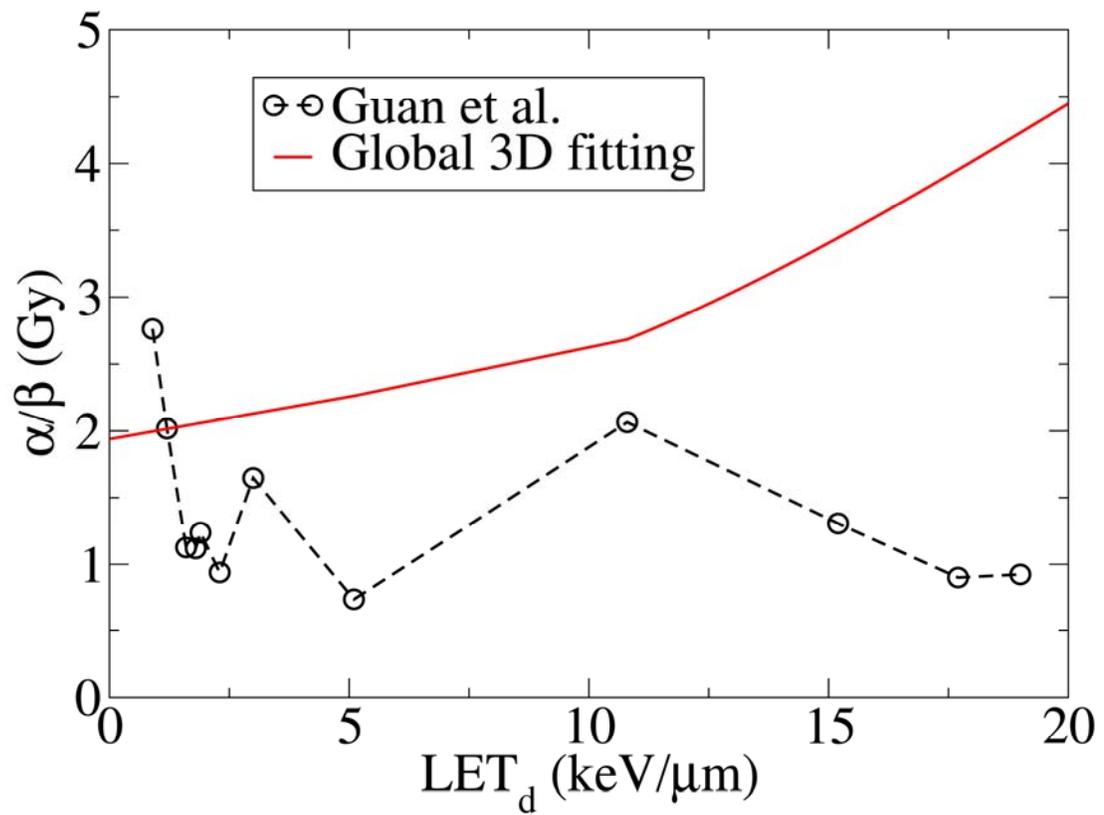

Figure 9: Lack of correlation among LETs in fitted cell-survival data in the conventional fitting procedure leads to strong fluctuations in $\alpha/\beta$ as reported by Guan et al. [37]. In contrast, $\alpha/\beta$ increases with increasing $LET_d$ in 3D global fitting consistent with the cell inactivation target theory (H460 cells).

|  | H460 | H460 | H1437 | H1437 |
|---|---|---|---|---|
|  | Low LET | High LET | Low LET | High LET |
| $b_{1,1}$ | -0.21447 | -0.12913 | -0.05237 | -0.12288 |
| $b_{2,1}$ | -0.007 | -6.733E-03 | -0.0151 | -0.00624 |
| $b_{3,1}$ | 0 | -4.563E-04 | 0 | -1.322E-06 |
| $b_{4,1}$ | 0 | -3.37529E-05 | 0 | -9.449E-07 |
| $b_{5,1}$ | 0 | -8.16229E-06 | 0 | -8.812E-07 |
| $b_{6,1}$ | 0 | 0 | 0 | -7.725E-08 |
| $b_{2,2}$ | -0.11061 | -0.06864 | -0.03265 | -0.00382 |
| $b_{3,2}$ | 0 | -0.00202 | 0 | -0.0007821 |
| $b_{4,2}$ | 0 | -1.36897E-04 | 0 | -9.275E-05 |
| $b_{5,2}$ | 0 | -3.63748E-05 | 0 | -9.511E-07 |
| $b_{6,2}$ | 0 | 0 | 0 | -1.68E-07 |
| Reduced $\chi^2$ | 0.00487 | 0.07557 | 0.00541 | 0.05472 |
| $R^2$ (COD) | 0.99496 | 0.98407 | 0.98233 | 0.96777 |
| Adjusted $R^2$ | 0.99481 | 0.92833 | 0.98233 | 0.93911 |

Table 1. Numerical values of expansion coefficients in $\alpha$ and $\beta$ given in Eqs. (3) and (4) for H460 and H1437 cell survivals. For a beam proton with energy 80 MeV the transition between low and high LETs occurs at $LET \approx 10 keV/\mu m$. The reduced chi-square, $\chi^2$, the coefficients of determination (COD), $R^2$, and adjusted $R^2$ have presented to illustrate the goodness of the fitting procedure.

**Conclusion**

We have developed an improved approach to fitting clonogenic survival data for therapeutic proton and ion beams. Our approach is based on three-parameter (3D) global fitting of the high-throughput, high accuracy survival fraction data acquired after irradiation with mono-energetic proton beams. In our approach, the intrinsic correlations in cell survival data are incorporated into the fitting algorithm. As a result, background fluctuations of the extracted parameters $\alpha$ and $\beta$ of the linear quadratic model due to uncertainties in the biological data are considerably reduced.

To demonstrate the reduction in numerical noise, we provided a comparison between traditional methods of post-processing of the clonogenic survival data and our approach. The numerical fluctuations in RBE parameters seen when an independent fitting of cell survival data is performed, can be suppressed by integrating the LET models including gMKM, LEM, and MCDS-RMF into 3D global fitting. The latter based on gMKM predicts a gradual increase in , $\beta$, and $\alpha/\beta$ as a function of LET, consistent qualitatively with the cell-inactivation target theory.

Further studies are on the way to implement other models such as multi-scale modeling [52, 53] as well as MCDS-RMF into 3D global fitting of the current clonogenic survival data (that non-linear LET terms in $\alpha$ and $\beta$ can be interpreted as higher order and more complex chromosome aberrations) as well as convolving the spectrum of lineal-energies, $d\varepsilon/dl$, calculated by MC on $\alpha$ and $\beta$ to take into account the effect of non-Poisson distribution. The latter is closely related to the ideas and approaches presented by Hawkins [29] and Kase *et al.* [54].

**Acknowledgement**

The authors would like to acknowledge useful discussion and scientific exchange with Drs., Jacek Capala, Alejandro Carabe, David Carlson, Fada Guan, Yusuf Helo, Armin Lühr, Dragan Mirkovic, Harald Paganetti, Darshana Patel, George Sandison, Robert Stewart, Reza

Taleei, Uwe Titt and Howard Thames. The work at the University of Texas, MD Anderson Cancer Center was supported by the NIH / NCI under Grant No. 2U19CA021239.

**Authors contributions:**
RA: wrote the main manuscript, appendix, prepared figures, performed mathematical derivations and computational steps including Geant4 and Geant4-DNA Monte Carlo simulations and three dimensional surface fitting to the experimental data. CRP & MN: wrote the manuscript and assisted with data analysis LB: provided experimental clonogenic cell survival data, aided in its interpretation and wrote the manuscript, DG & RM: wrote the main manuscript, proposed scientific problem and co-supervised the project.

**Competing financial interest:** The authors declare no competing financial interests.

**Supplementary Materials**

**Appendix – renormalization of radiobiological indices by energy loss fluctuations.**
This appendix describes the steps in the mathematical derivation of Eqs. (1) to (4) and the basis of the cell survival in the generalized microdosimetric kinetic model (gMKM) introduced in this study. More specifically we provide an analytical derivation of the solutions of the RMF model and the perturbative corrections around these solutions by taking into account the statistical fluctuations in energy deposition for an arbitrary distribution function, hence the renormalization of $\alpha$ and $\beta$. We consider a limiting case by systematically incorporating the contribution of the higher order fluctuations to the Gaussian distribution in energy deposition to correct the dependence of $\beta$ on lineal energy. Such perturbative expansion predicts an exact solution for $\alpha$ as a linear function of lineal energy (as known in the microdosimetry literature for decades), while $\beta$ shows quadratic dependence.

As pointed out in the main text, the linear density of the atomistic excitations and ionizations undergo a transition to a highly compact distribution as the primary protons slow down. Therefore, the deviation from a Poisson distribution appears to be significant at the end of the range of proton. Figures 1A and 1B in the main text illustrate the ionization events within the cellular dimensions generated by traversing a proton with initial energy of 80 MeV and 1 MeV. The MC simulation has performed by using Geant4 DNA. As illustrated in these figures, the number of events in low energies (e.g., 1 MeV) is orders of magnitude greater than high energies (e.g., 80 MeV). Such difference in compactness of the ionization clearly justifies deviation from the Poisson distribution as the proton keeps losing kinetic energy.

Refs. [28,29,49] describes the original approach in mathematical formulation of the cell survival in the standard microdosimetric kinetic model (MKM), starting from the repair-misrepair fixation model (RMF) [45-46] in a cell nucleus domain. In this approach, Hawkins has shown that the time-integrated solution of the linearized RMF mass-action equations, averaged over the ensemble of the cell nucleus domains, leads to the linear-quadratic dependence of cell survival on the deposited dose, the first two terms in Eq.(1). Moreover this model predicts $\alpha$ to be a linear function of LET and $\beta$ a constant and independent of LET.

A similar approach with a distinction of including the non-linear (quadratic) term in the solutions of the RMF mass-action equations that accounts for the chromosome misrepair binary end-joining leads to higher order deposited dose and LET terms. These terms in Eqs. (1) to (4) are effectively perturbative corrections to the linear expansion of $\alpha$ and $\beta$ calculated in the MK model as presented in Ref. [49].

**Cell survival in RMF and MK model:** The mass-action equations describing chromosome repair-misrepair binary end-joining introduced in Refs. [45-46] and [49] are given by

$$\frac{dn}{dt} = \mu \dot{z} - \lambda n - \gamma n^2, \qquad (A1)$$
$$\frac{dN}{dt} = \lambda_L n + \gamma_L n^2. \qquad (A2)$$

Here $n$ and $N$ represent the number of DSBs and the lethal lesions per cell in a domain, equivalent of type II and I lesions in the MK model [49], respectively. $\dot{z}$ is the microscopic dose rate, $\mu$ is the number of DSBs per domain per Gy. $\lambda$ and $\gamma$ are repair and binary misrepair end-joining coefficients corresponding to DSB restitution rate and binary DSB removal rate (the average rate at which binary misrepair removes DSBs by using them in lethal lesions or in harmless rearrangement), respectively.

Due to large fluctuations of energy deposition in sub-micrometer volumes, the ionizing radiation is characterized by the probability distribution of specific energy, and single event specific and lineal energies and their expectation values, the frequency-mean ($z_F, y_F$), and dose-averaged ($z_D, y_D$), as well as their higher order statistical moments. The cell survival fraction is therefore given by

$$SF = e^{-\bar{N}}, \qquad (A3)$$

where $\bar{N}$ is the mean lethal lesions, averaged over specific energy distribution, in the ensemble of domains in all cell nuclei.

**Linear solutions and LQ cell survival:** First we consider the linear approximation in RMF model, Eq. ($A3$), in which $\gamma = 0$, corresponding to Eq.(7) in Ref. [49]. In this limit, the analytical solution of Eqs. ($A1$) and ($A2$) can be easily found using the Green's function method

$$n_0(t) = \int_{-\infty}^{+\infty} dt' \, G_r(t - t') \dot{z}(t'). \qquad (A4)$$

Here $n_0$ is the solution of Eq. (A1) in the linear approximation. It is straightforward to justify that the homogeneous solution of Eq. (A1) is identical to zero, thus we do not consider it in Eq. (A4). It is also a straightforward calculation to show $G_r(t-t') = \mu e^{-\lambda(t-t')}\theta(t-t')$, where $\theta(t-t')$ is the Heavyside function, i.e., $\theta = 1$ if $t \geq t'$ and 0 otherwise. The steps in calculating retarded Green's function, $G_r$, include converting the integral equation, Eq. (A4), to a differential equation for $G$ by substituting (A4) in (A1) and imposing the initial condition $\bar{n}_0 = 0$ for $t < 0$ where $\dot{z} = 0$. Similarly we define $SF_0 = e^{-\bar{N}_0}$ where $\bar{N}_0 = \int_{-\infty}^{+\infty} dt' [\lambda_L \bar{n}_0 + \gamma_L \overline{n_0^2}]$. Here the bar over $N_0$ denotes energy deposition averaging on the ensemble of cell nuclei domains, specific to a lineal-energy distribution.

For an acute radiation dose, $\dot{z}(t) = z\delta(t)$, the solution of Eq.(A4), $n_0(t) = \mu z e^{-\lambda t}\theta(t)$, leads to $\bar{N}_0 = \frac{\lambda_L}{\lambda}\mu\bar{z} + \frac{\gamma_L}{2\lambda}\mu^2\overline{z^2}$. By averaging over the lineal-energy distribution and all cell nuclei and their domains we obtain a linear-quadratic model in cell-survival

$$-\ln(SF) = \alpha\bar{z} + \beta\bar{z}^2, \quad (A5)$$

where $\alpha = \frac{\lambda_L}{\lambda}\mu + \frac{\gamma_L}{2\lambda}\mu^2 z_D$ and $\beta = \frac{\gamma_L}{2\lambda}\mu^2$. Here we use the identity $\overline{z^2} = \bar{z}(\bar{z} + z_D)$ [50] that accounts for the spatial averaging of the energy deposition fluctuations, where $z_D = \frac{1}{z_F}\int_0^\infty z^2 f_1(z)dz$, $z_F = \int_0^\infty z f_1(z)\,dz$ and $f_1(z)$ is the single event distribution function of specific energy deposition, a counterpart distribution function of the lineal-energy $f(y)$. Furthermore $\bar{z} = \int_0^\infty zF(z; n_T)dz = n_T z_F$ where $F(z; n_T)$ is distribution function accounting for all events and $n_T$ is mean number of events and/or tracks. Applying a relation between $z$ and $y$ (see e.g., Eq.(II.28) in Ref. [49] or Eq. (8) in Ref. [54]), $z_D = l(y_D/m)$, where $m = \rho V$ and $l$ are the average mass and the chord length of a MKM domain, with $\rho$ and $V$, the mass density and the average volume of the domains, we obtain $\alpha = \frac{\lambda_L}{\lambda}\mu + \frac{\gamma_L}{2\lambda}\mu^2 \frac{1}{\rho(V/l)}y_D$. Considering the RMF and MKM constants in $\alpha$ and $\beta$ as phenomenological parameters we end up with two relations frequently used in the literature for fitting RBE data, (see for example Refs. [49,54])

$$\alpha = \alpha_0 + \beta \frac{1}{\rho\pi r_d^2}y_D; \quad \beta = \beta_x. \quad (A6)$$

Here $\alpha_0$, $\beta_x$, and $r_d$ (the radius of a spherical domain) are the phenomenological fitting parameters.

**Non-linear expansion of RMF solutions, going beyond LQ cell survival:** We now turn to perform a perturbative expansion to calculate the non-linear solution of the RMF model, Eqs. (A1) and (A2). To go beyond the RMF linear solutions presented above, we assume $\gamma$ to be a small parameter, hence we expand $n$ about $n_0$ perturbatively and linearize the resulting mass-action kinetic equation to obtain the dynamics of the small fluctuations describing deviations from linear DSB solutions. We define $n_1 = n - n_0$ and recall Eq. (A1) to obtain a linear mass-action equation for $n_1$

$$\frac{dn_1}{dt} = -\lambda n_1 - \gamma(2n_0 n_1 + n_0^2) + O(n_1^2). \quad (A5)$$

Here $n$ is the exact solution of the RMF model. As we defined, $n_0$ is the linear solution of the RMF model, hence $n_1$ describes the difference between exact and linear solutions. It is more convenient to transform Eq. (A5) into a more compact form

$$\frac{dn_1}{dt} + \eta(t)n_1(t) = \xi(t), \quad (A6)$$

where $\eta = \lambda + 2\gamma n_0$ and $\xi = \gamma n_0^2$. The solution of Eq. (A6) can be calculated exactly

$$n_1(t) = e^{-\varphi(t)} \int_{-\infty}^{t} dt' \xi(t') e^{\varphi(t')}, \quad (A7)$$

where $\varphi(t) = \lambda t + 2\gamma \int_{-\infty}^{t} dt' n_0(t')$. Linearizing Eq. (A7) in terms of $\gamma$, assuming $\gamma$ is a small parameter, leads to

$$n_1(t) = \gamma e^{-\lambda t} \int_{-\infty}^{t} dt' n_0^2(t') e^{\lambda t'} + O\left(\frac{\gamma^2}{\lambda^2}\right). \quad (A8)$$

Substituting the linear solution calculated above, $n_0(t) = \mu z e^{-\lambda t} \theta(t)$, in Eq. (A8) yields

$$n_1(t) = \frac{\gamma}{\lambda} \mu^2 z^2 (1 - e^{-\lambda t}) e^{-\lambda t}, \quad (A9)$$

hence

$$n = n_0 + n_1 = n_0 - \frac{\gamma}{\lambda}[n_0 - \mu z] n_0 + O(n_0^3). \quad (A10)$$

From Eq.(A10) and $n_0$ the cell-survival can be calculated, $-\ln(\text{SF}) = \int_{-\infty}^{+\infty} dt' \left[\lambda_L \bar{n} + \gamma_L \overline{n^2}\right]$, hence

$$-\ln(\text{SF}) = \frac{\lambda_L}{\lambda} \mu \bar{z} + \frac{1}{2}\left[\frac{\lambda_L}{\lambda}\frac{\gamma}{\lambda} + \frac{\gamma_L}{\lambda}\right]\mu^2 \overline{z^2} + \frac{1}{3}\frac{\gamma_L}{\lambda}\frac{\gamma}{\lambda}\mu^3 \overline{z^3} + \left[-\frac{\gamma_L}{6\lambda}\frac{\gamma^2}{\lambda^2}\mu^4 \overline{z^4} + O\left(\frac{\gamma^2}{\lambda^2}\overline{z^4}\right)\right]$$
$$+ O(\overline{z^5}) \quad (A11)$$

The last two terms in Eq. (A11) are the contribution of the terms omitted in Eq. (A8) due to linearizing $n$ in the limit where $\gamma$ is negligible. To transform (A11) to a form similar to the linear-quadratic model we must calculate the statistical fluctuations in microscopic dose deposition throughout the averaging over cell nucleus domains, assuming equivalence between the ensemble averaging over the domains and the spatial averaging of the energy deposition fluctuations over the cell nuclei. In general, these fluctuations can be recursively reduced to lower power fluctuations, namely

$$\overline{z^2} = \bar{z}^2 + z_{21}\bar{z}, \quad (A12a)$$
$$\overline{z^3} = \bar{z}^3 + z_{32}\bar{z}^2 + z_{31}^2 \bar{z}, \quad (A12b)$$
$$\overline{z^4} = \bar{z}^4 + z_{43}\bar{z}^3 + z_{42}^2 \bar{z}^2 + z_{41}^3 \bar{z}. \quad (A12c)$$

...

and in general

$$\overline{z^i} = \sum_{j=1}^{i} z_{ij}^{i-j} \bar{z}^j$$

Here $z_{ij}$ are coefficients in the expansion with the physical dimension identical to the dimension of specific energy, Gy, and can be calculated by integrating over single event energy deposition distribution, $f_1(z)$. For example $z_{21} = z_D$, $z_{31}^2 = \frac{1}{z_F}\int_0^\infty z^3 f_1(z)dz$, $z_{32} = 3z_D$, $z_{41}^3 = \frac{1}{z_F}\int_0^\infty z^4 f_1(z)dz$, $z_{42}^2 = 3z_D^2 + 4z_{31}^2$, $z_{43} = 6z_D$, …

Applying Eqs. (A12a) to (A12c) in Eq. (A11) and keeping up to the quadratic term in macroscopic dose, $D = \bar{z}$, we find

$$-\ln(\text{SF}) = \left[\frac{\lambda_L}{\lambda}\mu + \frac{1}{2}\left(\frac{\lambda_L}{\lambda}\frac{\gamma}{\lambda} + \frac{\gamma_L}{\lambda}\right)\mu^2 z_{21} + \frac{1}{3}\frac{\gamma_L}{\lambda}\frac{\gamma}{\lambda}\mu^3 z_{31}^2 + \cdots\right]\bar{z}$$
$$+ \left[\frac{1}{2}\left(\frac{\lambda_L}{\lambda}\frac{\gamma}{\lambda} + \frac{\gamma_L}{\lambda}\right)\mu^2 + \frac{1}{3}\frac{\gamma_L}{\lambda}\frac{\gamma}{\lambda}\mu^3 z_{32} + \cdots\right]\bar{z}^2 + \cdots \quad (A13)$$

Hence

$$\alpha = \frac{\lambda_L}{\lambda}\mu + \frac{1}{2}\left(\frac{\lambda_L}{\lambda}\frac{\gamma}{\lambda} + \frac{\gamma_L}{\lambda}\right)\mu^2 z_{21} + \frac{1}{3}\frac{\gamma_L}{\lambda}\frac{\gamma}{\lambda}\mu^3 z_{31}^2 + \cdots \quad (A14)$$

$$\beta = \frac{1}{2}\left(\frac{\lambda_L}{\lambda}\frac{\gamma}{\lambda} + \frac{\gamma_L}{\lambda}\right)\mu^2 + \frac{1}{3}\frac{\gamma_L}{\lambda}\frac{\gamma}{\lambda}\mu^3 z_{32} + \cdots \quad (A15)$$

In the limit where $\gamma = 0$, Eqs. (A14) and (A15) reduce to $\alpha = \frac{\lambda_L}{\lambda}\mu + \frac{\gamma_L}{2\lambda}\mu^2 z_D$ and $\beta = \frac{\gamma_L}{2\lambda}\mu^2$, the linear MKM approximation [46]. By expanding $z_{ij}$ around $z_D$ in Eqs. (A14) and (A15), e.g., $\int_0^\infty z^n f_1(z)dz = \int_0^\infty (z_D + \delta z)^n f_1(z)dz = \sum_{k=0}^n \frac{n!}{k!(n-k)!} z_D^k \int_0^\infty (z - z_D)^{n-k} f_1(z)dz = \sum_{k=0}^n \frac{n!}{k!(n-k)!} z_D^k \tilde{f}_{n-k}(z_D)$ where $\delta z = z - z_D$ and $\tilde{f}_k(z_D) = \int_0^\infty (z - z_D)^k f_1(z)dz$, and using $z_{ij} = ly_{ij}/m$, and a linear relationship between $y_D$ and $L$ [55,56], one may find $\alpha$ and $\beta$ to be a power series in lineal energy and LET as given by Eq.(3) and (4). The constants in Eqs. (A14) and (A15) can be determined phenomenologically by fitting $\alpha$ and $\beta$ to the experimental data as illustrated in the text, i.e., $\alpha = \sum_{k=1}^n b_{k,1} L^{k-1}$, and $\beta = \sum_{k=1}^{n-1} b_{k+1,2} L^{k-1}$. As seen in these equations, the contribution from the energy loss fluctuations renormalizes the LQ biological parameters $\alpha$ and $\beta$ to infinite orders in $z_D$ and $y_D$.

**Gaussian fluctuations**
It is interesting to calculate $z_{ij}$, $\alpha$ and $\beta$ for a widely used Gaussian distributed function as in the limit $\bar{z} \gg 0$, the Poisson distribution can be approximated to a Gaussian (central limit theorem) with variance $\sigma^2 = z_{GD}\bar{z}_G$

$$F(z; n_T) \rightarrow F_G(z; n_T) = \frac{1}{\sigma\sqrt{2\pi}}\exp\left(-\frac{(z - \bar{z}_G)^2}{2\sigma^2}\right)$$

Here we change the notations $\rightarrow F_G$, $\bar{z} \rightarrow \bar{z}_G$, $z_F \rightarrow z_{GF}$ and $z_D \rightarrow z_{GD}$ with the subscript $G$ to denote the averaging over the Gaussian distribution function. It is straightforward calculation to find the Gaussian version of equations (A12a – A12c)

$$\overline{z^2} \rightarrow \overline{z_G^2} = \bar{z}_G^2 + z_{GD}\bar{z}_G, \quad (A12d)$$
$$\overline{z^3} \rightarrow \overline{z_G^3} = \bar{z}_G^3 + 3z_{GD}\bar{z}_G^2, \quad (A12e)$$
$$\overline{z^4} \rightarrow \overline{z_G^4} = \bar{z}_G^4 + 6z_{GD}\bar{z}_G^3 + 3z_{GD}^2\bar{z}_G^2, \quad (A12f)$$
$$\overline{z^5} \rightarrow \overline{z_G^5} = \bar{z}_G^5 + 10z_{GD}\bar{z}_G^4 + 15z_{GD}^2\bar{z}_G^3, \quad (A12g)$$
$$\overline{z^6} \rightarrow \overline{z_G^6} = \bar{z}_G^6 + 15z_{GD}\bar{z}_G^5 + 45z_{GD}^2\bar{z}_G^4 + 15z_{GD}^3\bar{z}_G^3, \quad (A12h)$$

where all statistical fluctuations and higher order moments are reduced to two variables $z_{GD} = \frac{1}{z_{GF}} \int_0^\infty z^2 f_{G1}(z) dz$ and $\bar{z}_G = n_T z_{GF}$ where $z_{GF} = \int_0^\infty z f_{G1}(z) dz$. A comparison between Eqs. (A12d – A12f) and (A12a – A12c) shows that $z_{G,21} = z_D, z_{G,32} = 3z_D, z_{G,31} = 0, z_{G,43} = 6z_D, z_{G,42} = \sqrt{3}z_D, z_{G,41} = 0, \ldots$ Note that the fluctuations such as $z_{G,31}, z_{G,32}, z_{G,41}$ are negligible because of specific symmetry of $G(z)$.

Similar to a general energy loss fluctuations, the contributions of the Gaussian fluctuations in energy deposition introduce correction factors to $\alpha$ and $\beta$. Because the contribution to $\bar{z}_G$ from $\overline{z_G^3}$ and beyond are identical to zero, there is no correction beyond the linear term in $z_{GD}$ and hence in $y_{GD}$. This is evident from Eq. $(A12e)$ as the lowest order correction to specific energy and deposition dose is quadratic. Hence the Gaussian model predicts $\alpha$ to be only linear dependence on lineal energy and LET

$$\alpha_G = \frac{\lambda_L}{\lambda}\mu + \frac{1}{2}\left(\frac{\lambda_L}{\lambda}\frac{\gamma}{\lambda} + \frac{\gamma_L}{\lambda}\right)\mu^2 z_{GD}. \qquad (A16)$$

Similarly for $\beta$ the corrections beyond $\overline{z^5}$ fluctuations vanish exactly as the lowest order correction to the dose in Eq. $(A12g)$ is cubic. Hence we find a closed form for $\beta$

$$\beta_G = \frac{1}{2}\left(\frac{\lambda_L}{\lambda}\frac{\gamma}{\lambda} + \frac{\gamma_L}{\lambda}\right)\mu^2 + \frac{\gamma_L}{\lambda}\frac{\gamma}{\lambda}\mu^3 z_{GD} + \left(-\frac{\gamma_L}{6\lambda}\frac{\gamma^2}{\lambda^2} + O\left(\frac{\gamma^2}{\mu^4 \lambda^2}\right)\right) 3\mu^4 z_{GD}^2. \qquad (A17)$$

As seen in these equations, the contribution from the fluctuations in Gaussian model of energy deposition correct the LQ biological parameters $\alpha_G$ and $\beta_G$ to scale linearly and quadratically in $z_D$ and $y_D$. We note that the radio-biological models that start from the equations equivalent of $(A16)$ and $(A17)$ are implicitly assuming a Gaussian-type symmetry in the energy-loss processes. This includes models with linear dependence on LET in $\alpha$ and constant $\beta$ (e.g., by neglecting higher order $\mu$ terms in $\alpha$ and $\beta$ beyond quadratic terms). Such models neglect the importance of asymmetries hidden in more realistic distribution functions such as the Landau [57] and/or Vavilov [58] distribution functions, responsible for observed non-linearities in biological responses.

**Neyman's distribution of type A and DSB distribution**

We devote this section to illustrate the effect of deviation from Poisson distribution on statistical fluctuations of specific energy, DSB distribution, and biological indices $\alpha$ and $\beta$. Specifically we start with construction of the Neyman's distribution function from first principles for DSB induction processes and calculate the higher order moments in specific energy needed for incorporation of the repair and mis-repair mechanisms to fit globally the cell survival data.

We consider the normalized distribution function to describe stochastic energy deposition in DNA material [50,55] as given below

$$F(z; \bar{v}) = \sum_{v=0}^{\infty} p_v(\bar{v}) f_v(z), \qquad (A18)$$

where $p_v(\bar{v}) = (\bar{v}^v/v!)\exp(-\bar{v})$ describes the Poisson distributed events in an ensemble of single tracks. Here $\bar{v}$ denotes the average number of energy deposition events and $f_v(z)$ is the distribution of specific energy imparted from passage of a single track within $z$ and $z + dz$ resulted from exactly $v$ energy deposition events. The stochastic process as such is sketched in Fig. (1A).

We denote $\varepsilon_\nu$ the deposited energy resulted from exactly $\nu$ events in mass $m$ of DNA material corresponding to specific energy $z = \frac{1}{m}\sum_{i=1}^{\nu}\varepsilon_i = \nu\bar{\varepsilon}_\nu/m$ where $\bar{\varepsilon}_\nu = \frac{1}{\nu}\sum_{i=1}^{\nu}\varepsilon_i$. The occurrence of $k$ DSBs resulted from energy deposition requires balance in energy transfer to DNA. More specifically $\nu\bar{\varepsilon}_\nu = \nu\sum_{i=1}^{\Delta}\epsilon_i = k\bar{\epsilon}_\Delta$ where $\bar{\epsilon}_\Delta = \frac{1}{\Delta}\sum_{i=1}^{\Delta}\epsilon_i$ is the typical energy results in breaking chemical bonds for induction of a single DSB and $k = \nu\Delta = 0,1,2,\ldots$ counts number of DSBs. $\Delta$ is the number of induced DSBs in an event. Hence the energy balance in an exactly $\nu$ events processes requires $z = \bar{\epsilon}_\Delta \nu\Delta/m$. Further simplification can be performed by averaging over DSB population per event subsequently by averaging over events that yields $\bar{\bar{k}} = \overline{\nu\Delta} = \bar{\nu}\bar{\Delta}$. Here $\bar{\Delta}$ is average number of DSBs per event, independent of $\nu$. The double bars over $k$ denotes two independent averagings; hence the order of averaging is not an issue. Also note that $n = \bar{\bar{k}}$ and $\mu = m/\bar{\epsilon}_\Delta$ in Eq.(A1) hence $\dot{z} = (\bar{\Delta}/\mu)\dot{\bar{\nu}}$.

The distribution of DSBs in a class of events, specified by given $\nu$, can be uniquely determined by the corresponding energy deposition distribution. The DSB partition function, $Q_k(\bar{\Delta};\bar{\nu})$, the probability distribution in finding exactly $k$ DSBs, can be calculated from Eq.(A18) where $f_\nu(z) = \sum_{k=0}^{\infty}p_k(z)\delta(z - \nu\bar{\Delta}/\mu)$. The insertion of $\delta$-function, the DSB density of states, enforces a constraint on the energy transfer balance resembling Fermi golden rule formulation of transition rates and the perturbation theory in quantum physics. Substituting $f_\nu(z)$ in Eq.(A18) and integrating over $z$ yields

$$1 = \int_0^\infty F(z;\bar{\nu})\,dz = \sum_{\nu=0}^{\infty}p_\nu(\bar{\nu})\int_0^\infty \sum_{k=0}^{\infty}p_k(z)\delta(z - \nu\bar{\Delta}/\mu)\,dz$$

$$= \sum_{k=0}^{\infty}\sum_{\nu=0}^{\infty}p_\nu(\bar{\nu})p_k(\nu\bar{\Delta}) = \sum_{k=0}^{\infty}Q_k(\bar{\Delta};\bar{\nu}), \qquad (A19)$$

Thus

$$Q_k(\bar{\Delta};\bar{\nu}) = \sum_{\nu=0}^{\infty}p_\nu(\bar{\nu})p_k(\nu\bar{\Delta}). \qquad (A20)$$

Further approximation can be performed by considering Poisson distribution for DSB events, i.e., $p_k(\nu\bar{\Delta}) = (\nu\bar{\Delta})^k/k!\exp(-\nu\bar{\Delta})$ that reduces $Q_k(\bar{\Delta};\bar{\nu})$ in Eq.(A20) to Neyman's distribution of Type A (see for example Refs. [57,58]). Accordingly, the probability in finding a DNA with zero DSB is given by $Q_0 = \exp(-\bar{\nu}(1 - \exp(-\bar{\Delta}))) \approx \exp(-\bar{\nu}\bar{\Delta})$. In this equation, $\bar{\nu}\bar{\Delta}$ is the average number of events times the average number of DSBs per event.

From Eq. (A20), it is now straightforward to calculate the statistical moments of DSBs and the corrections to biological indices $\alpha$ and $\beta$ by inclusion of DSB fluctuations, as discussed in the main text. Here we systematically show that the DSB partition function given by $Q_k(\bar{\Delta};\bar{\nu})$ in Eq. (A20) provides all statistical moments we need for this analysis. For clarification of notations we first check the self-consistency of equations by calculating the first moment

$$\bar{\bar{k}} = \sum_{k=0}^{\infty}kQ_k(\bar{\Delta};\bar{\nu}) = \bar{\nu}\bar{\Delta}.$$

Furthermore

$$\overline{k(k-1)} = \sum_{k=0}^{\infty}k(k-1)Q_k(\bar{\Delta};\bar{\nu}) = (\bar{\nu}\bar{\Delta})^2 + (\bar{\nu}\bar{\Delta})\bar{\Delta} = \bar{\bar{k}}^2 + \bar{\bar{k}}\bar{\Delta},$$

$$\overline{k(k-1)(k-2)} = \sum_{k=0}^{\infty}k(k-1)(k-2)Q_k(\bar{\Delta};\bar{\nu}) = \bar{\bar{k}}^3 + 3\bar{\bar{k}}^2\bar{\Delta} + \bar{\bar{k}}\bar{\Delta}^2,$$

and in general

$$\overline{k(k-1)(k-2)\ldots(k-r)} = \sum_{s=0}^{r} c_s \bar{k}^{r-s+1}\overline{\Delta}^s. \qquad (A21)$$

Here $c_s$ are the coefficients of expansion and $c_0 = 1$. By further expansion of Eq. (A21) we find a power law series dependence of DSB fluctuations on $\Delta$ that we need for the expansion of Eqs. (A1) and (A2)

$$\overline{\overline{k^r}} = \bar{k}^r + (a_{01} + a_{11}\overline{\Delta})\bar{k}^{r-1} + (a_{02} + a_{12}\overline{\Delta} + a_{22}\overline{\Delta}^2)\bar{k}^{r-2} + \cdots$$
$$+ (a_{0r} + a_{1r}\overline{\Delta} + a_{2r}\overline{\Delta}^2 + \cdots + a_{rr}\overline{\Delta}^r)\bar{\bar{k}}. \qquad (A22)$$

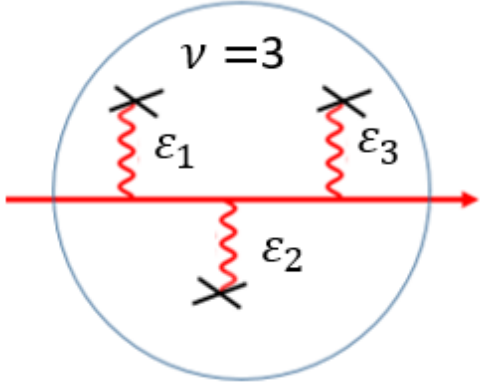

**Figure 1A.** Schematically shown the energy deposition of $\nu = 3$ events resulted from passage of a single track in a cell domain.

Finally we show that $\overline{\Delta} = \mu z_D$. This identity is the reminiscent of fluctuation-dissipation theorem and can be derived in the following steps

$$\frac{\overline{z^2}}{\bar{z}} = \frac{\int_0^\infty z^2 F(z;\bar{\nu})dz}{\int_0^\infty z F(z;\bar{\nu})dz} = \frac{\sum_{\nu=0}^\infty p_\nu(\bar{\nu})\int_0^\infty z^2 dz \sum_{k=0}^\infty p_k(z)\delta\left(z - \frac{\nu\overline{\Delta}}{\mu}\right)}{\sum_{\nu=0}^\infty p_\nu(\bar{\nu})\int_0^\infty z\, dz \sum_{k=0}^\infty p_k(z)\delta\left(z - \frac{\nu\overline{\Delta}}{\mu}\right)}.$$

After integrating over $\delta$-function we find

$$\frac{\overline{z^2}}{\bar{z}} = \frac{\sum_{k=0}^\infty \sum_{\nu=0}^\infty p_\nu(\bar{\nu})p_k(\nu\overline{\Delta})(\nu\overline{\Delta}/\mu)^2}{\sum_{k=0}^\infty \sum_{\nu=0}^\infty p_\nu(\bar{\nu})p_k(\nu\overline{\Delta})(\nu\overline{\Delta}/\mu)} = \frac{\overline{\Delta}\sum_{\nu=0}^\infty \nu^2 p_\nu(\bar{\nu})\sum_{k=0}^\infty p_k(\nu\overline{\Delta})}{\mu \sum_{\nu=0}^\infty \nu p_\nu(\bar{\nu})\sum_{k=0}^\infty p_k(\nu\overline{\Delta})}.$$

Because $\sum_{k=0}^\infty p_k(\nu\overline{\Delta}) = 1$ and $\sum_{\nu=0}^\infty \nu^2 p_\nu(\bar{\nu}) = \overline{\nu^2} = \bar{\nu}(\bar{\nu}+1)$ and $\sum_{\nu=0}^\infty \nu p_\nu(\bar{\nu}) = \bar{\nu}$ we find

$$\frac{\overline{z^2}}{\bar{z}} = \frac{\overline{\Delta}}{\mu}(\bar{\nu}+1),$$

hence $\overline{\Delta} = [\mu/(\bar{\nu}+1)](\overline{z^2}/\bar{z})$. For a class of single events, $\overline{z^2}/\bar{z}$ reduces to $z_D$. More specifically

$$z_D = \frac{\overline{z^2}}{\bar{z}}\bigg|_{\nu=1} = \frac{\int_0^\infty z^2 F(z;\bar{\nu})dz}{\int_0^\infty z F(z;\bar{\nu})dz}\bigg|_{\nu=1} = \frac{\sum_{\nu=0}^\infty p_\nu(\bar{\nu})\int_0^\infty z^2 dz \sum_{k=0}^\infty p_k(z)\delta\left(z - \frac{\nu\overline{\Delta}}{\mu}\right)\delta_{\nu,1}}{\sum_{\nu=0}^\infty p_\nu(\bar{\nu})\int_0^\infty z\, dz \sum_{k=0}^\infty p_k(z)\delta\left(z - \frac{\nu\overline{\Delta}}{\mu}\right)\delta_{\nu,1}}.$$

Similarly after integrating over $\delta$-function we find

$$z_D = \frac{\sum_{k=0}^\infty \sum_{\nu=0}^\infty p_\nu(\bar{\nu})p_k(\nu\overline{\Delta})(\nu\overline{\Delta}/\mu)^2 \delta_{\nu,1}}{\sum_{k=0}^\infty \sum_{\nu=0}^\infty p_\nu(\bar{\nu})p_k(\nu\overline{\Delta})(\nu\overline{\Delta}/\mu)\delta_{\nu,1}} = \frac{\overline{\Delta}\sum_{\nu=0}^\infty \nu^2 p_\nu(\bar{\nu})\sum_{k=0}^\infty p_k(\nu\overline{\Delta})\delta_{\nu,1}}{\mu \sum_{\nu=0}^\infty \nu p_\nu(\bar{\nu})\sum_{k=0}^\infty p_k(\nu\overline{\Delta})\delta_{\nu,1}} = \frac{\overline{\Delta}}{\mu}.$$

Combining the results of these equations into Eq.(A22) and using transformation, $k \to \mu z$ in Eq.(A22) and multiplying that equation by $\mu^{-r}$ and substituting $\bar{\Delta} = \mu z_D$, we obtain

$$\overline{z^r} = \bar{z}^r + (b_{01} + b_{11}z_D)\bar{z}^{r-1} + (b_{02} + b_{12}z_D + b_{22}z_D^2)\bar{z}^{r-2} + \cdots$$
$$+ (b_{0r} + b_{1r}z_D + b_{2r}z_D^2 + \cdots + b_{rr}z_D^r)\bar{z}. \quad (A32)$$

Here $b_{sr} = a_{sr}\mu^{s-r}$ are the expansion coefficients. In numerical fitting procedure to the experimental cell-survival data we consider these coefficients as phenomenological adjustable parameters. Recalling $z_D = l(y_D/m)$ and following the derivation steps presented in previous section, we arrive at similar expressions for $\alpha$ and $\beta$ and the power series dependences on lineal energy.

The results given in Eqs. (A19-A22) for Neyman's distribution is in contrast with the Poisson distribution that only leads to the first term in the expansion given by Eq. ($A21$) and (A22)

$$\overline{k(k-1)(k-2)\ldots(k-r)} = \bar{k}^r \quad (A23)$$

Hence

$$\overline{\overline{k^r}} = \bar{\bar{k}}^r + a_{01}\bar{\bar{k}}^{r-1} + a_{02}\bar{\bar{k}}^{r-2} + \cdots + a_{0r}\bar{\bar{k}}, \quad (A24)$$

with no dependence on $\Delta$ and lineal energy in Eq. (A24). Thus

$$\overline{z^r} = \bar{z}^r + b_{01}\bar{z}^{r-1} + b_{02}\bar{z}^{r-2} + \cdots + b_{0r}\bar{z}. \quad (A25)$$

Substituting these results into Eqs. (A12a-c), it is now straightforward to show that the Poisson distribution leads to no dependence of $\alpha$ and $\beta$ on lineal energy and LET consistent with the results presented by Sachs *et al.* [46]. In contrast the Neyman's distribution leads to a power law dependence of $\alpha$ and $\beta$ on lineal energy and LET as discussed in the main text. The use of Neyman's distribution in RMF-MCDS model reported by Carlson *et al.* in Ref. [33] where $z_F$ and LET are resulted linearly in $\alpha$ where $\beta$ shows no dependence. One difference between derivation presented in our current study and the one presented in Ref. [33] is the incorporation of fluctuations that supersede $z_D$, instead of $z_F$ in $\alpha$ and $\beta$ such that the formulation of our linear model shows consistency with Hawkins' MK linear models [28,29,49].